\documentclass{article}
\usepackage{spconf,amsmath,graphicx,hyperref, booktabs, comment}
\usepackage{siunitx}
\usepackage{multirow}
\usepackage{orcidlink}

\newcommand{\pmci}[1]{{\scriptsize$\pm$#1}}

\makeatletter
\def\title#1{\gdef\@title{\MakeUppercase{#1}}}
\def\name#1{\gdef\@name{\shortstack[c]{\itshape #1\\[0.5em]}}}
\makeatother

\title{Adapting Appearance-Based Gaze Estimation to Narrow-Range, Long-Duration Screen Viewing}
\name{Jordan Prescott$^{*1}$\,\orcidlink{0009-0002-0914-4150},\quad Kleanthis Avramidis$^{*1}$\,\orcidlink{0000-0003-0308-795X},\quad Shrikanth Narayanan$^{1}$\,\orcidlink{0000-0002-1052-6204}
\thanks{\noindent $^*$Equal contribution.}
\thanks{This study was supported in part by DARPA under cooperative agreement No. N660012324006 and in part by ARPA-H. The content of the information does not necessarily reflect the position or the policy of the Government, and no official endorsement should be inferred.}
}
\address{$^1$ Signal Analysis and Interpretation Lab, University of Southern California}

\begin{document}
\ninept
\setlength{\abovedisplayskip}{3pt plus 1pt minus 1pt}
\setlength{\belowdisplayskip}{3pt plus 1pt minus 1pt}
\setlength{\abovedisplayshortskip}{2pt plus 1pt minus 1pt}
\setlength{\belowdisplayshortskip}{2pt plus 1pt minus 1pt}

\maketitle
\begin{abstract}
Appearance-based gaze estimation offers a low-cost alternative to infrared eye tracking for screen-based behavioral and clinical applications, however existing models are typically developed for wide ranges of gaze angle and head pose. Prolonged screen viewing presents a distinct regime in which gaze remains near the screen center, head motion is limited, and calibration drift accumulates over time. In this work, we benchmark six published estimators along with a proposed differential-gaze model on 140 long-duration facial video recordings with synchronized eye tracking under subject-disjoint evaluation and a common budget of calibration frames. Differential estimation was the only static method that significantly improved upon a baseline predictor of each recording's mean gaze, and was further boosted by addition of temporal context. It also yielded the strongest agreement with reference fixations and saccades, demonstrating that low angular error alone could not validate eye movement reconstruction. These findings establish differential estimation as a promising foundation for reliable gaze tracking in long-duration, narrow-range settings.
\end{abstract}

\section{Introduction}
\label{sec:intro}

Appearance-based gaze estimation, in which eye movements are inferred from camera recordings of the face, has been adopted as a low-cost and unobtrusive alternative to infrared eye tracking for screen-based behavioral and clinical analysis~\cite{papoutsaki2015scalable,semmelmann2018online}, such as for example in studies of reading~\cite{hutt2024webcam} and visual attention~\cite{xu2015turkergaze}. The most effective current systems are based on convolutional and transformer models trained on curated datasets that span wide ranges of head pose and gaze angle~\cite{kellnhofer2019gaze360,zhang2020ethxgaze,fischer2018rtgene,zhang2017fullface,cheng2022gaze}. Model accuracy for this task is typically reported as the mean angular error over those ranges.

Many clinical studies that would benefit from appearance-based gaze estimation involve prolonged screen viewing~\cite{hutt2024webcam, xu2015turkergaze}, where the gaze range is narrow and concentrated near the screen center, the head is near-stationary, and the sessions extend long enough for the eye-tracker calibration to drift~\cite{precog-methods}. Such conditions are largely absent from the data on which current gaze estimation models are trained~\cite{kellnhofer2019gaze360,zhang2020ethxgaze}, and hence their robustness in low-resource settings has been little examined. Moreover, the constant predictor, i.e., the per-recording mean gaze, is already accurate in narrow ranges, so a low mean angular error is not necessarily evidence of model robustness~\cite{chen2020offset}. As a result, more descriptive metrics should be evaluated.

In this study we investigate this regime as a case study on paired data of synchronized facial video and high-fidelity eye tracking as ground truth, from a long-duration screen-viewing task. Published appearance-based gaze models, combined with a lightweight subject adaptation, were evaluated under a common subject-independent protocol, together with behaviorally-grounded metrics defined at the level of fixations and saccades. A custom model based on differential gaze estimation was the only frame-wise system to significantly outperform the constant predictor, while significantly improving gaze metrics. Our contributions are summarized as follows:

\begin{itemize}
    \item We introduce a benchmark for appearance-based gaze estimation under narrow-range screen viewing, built on synchronized facial video and high-fidelity eye tracking.
    \item We also contribute a suite of behaviorally-grounded evaluation metrics including per-recording fixation and saccade-level angle agreement to complement angular error.
    \item We show that published models, evaluated under a subject-independent protocol, barely exceed the per-recording constant predictor after anchor-based calibration.
    \item We demonstrate that a model trained on differential gaze estimation and conditioned on temporal context yields substantial improvements in gaze metrics of narrow-range viewing.
\end{itemize}

\section{Dataset}
\label{sec:dataset}

\begin{figure*}[t]
\centering
\includegraphics[width=0.95\textwidth, height=0.24\textheight]{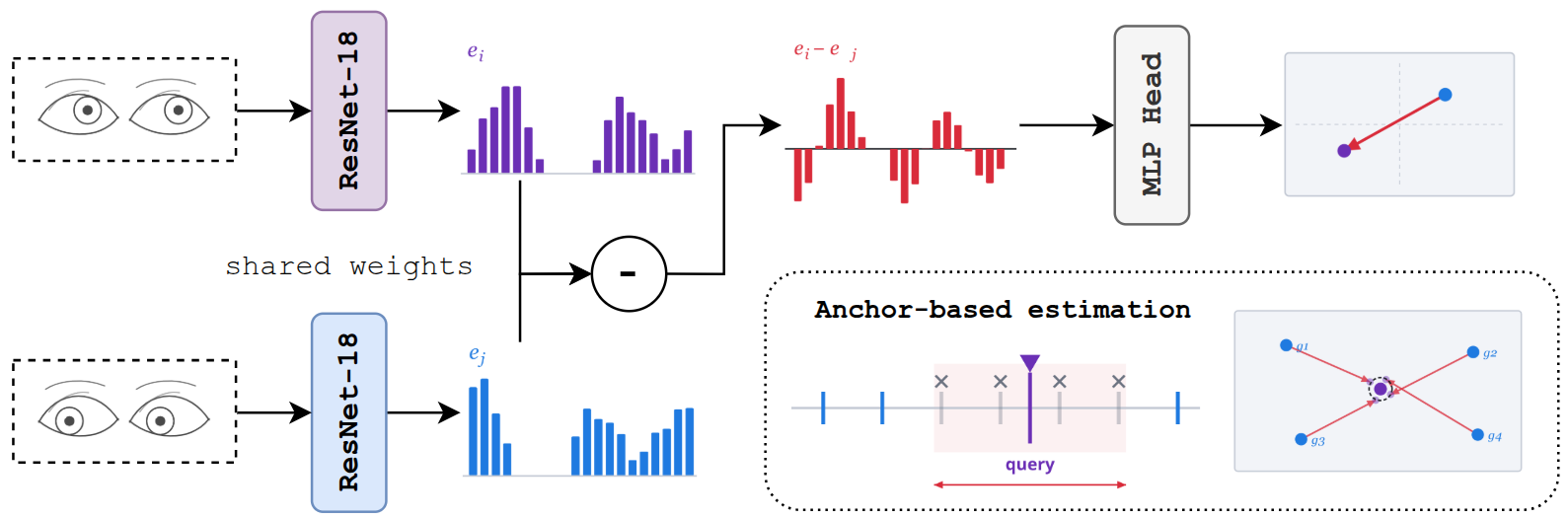}
\vspace{-2mm}
\caption{\textbf{Differential gaze estimation}. Two eye crops from the same recording are encoded by a weight-shared ResNet-18 into embeddings $e_i$ and $e_j$, and a bias-free MLP head maps their difference to a two-dimensional gaze displacement, drawn at right as the vector between the two gaze points. Absolute gaze is recovered by adding the predicted displacement to a labeled anchor from the same recording.}
\label{fig:method}
\vspace{-4mm}
\end{figure*}

Our gaze estimation benchmark uses paired facial video and high-fidelity eye tracking recordings from the PRECOG project~\cite{precog-methods, avramidis2026deep}, a multimodal study investigating the neural and cognitive mechanisms underlying depression and suicidal ideation. The original data collection included 146 college-aged adults who completed two cognitive tasks while neural, physiological and behavioral signals were recorded~\cite{precog-methods}. For the purpose of this study we use facial video acquired at 195\,Hz with a monochrome Teledyne FLIR Blackfly S camera and gaze data acquired at 250\,Hz with an EyeLink 1000 Plus tracker. The resulting dataset contains an average of 75 minutes of paired recordings per participant, totaling 190 hours.

Both tasks were administered using the same camera configuration and a 1920 $\times$ 1080 display measuring 510 $\times$ 286\,mm, viewed from a distance of 862.7\,mm and subtending 33\textdegree\ horizontally and 18.8\textdegree\ vertically. Participants completed either a fast-paced Stroop task comprising 480 trials interleaved with brief rest periods or a longer sentence-processing task interleaved with free-viewing intervals~\cite{precog-methods}. Despite differing in cognitive demand, both tasks exhibit the challenging viewing regime considered here: gaze remained predominantly near the screen center, head movement was limited, and sessions lasted long enough for participant pose and eye-tracker calibration to potentially drift. Annotations from the EyeLink eye tracker were additionally used as reference intervals for model estimates of fixations, saccades, and blinks.


\section{Methods}
\label{sec:methods}
\vspace{-2mm}

\subsection{Benchmark Models}

To evaluate appearance-based gaze estimation in our setting we benchmarked six published model configurations spanning full-face and eye-region inputs, convolutional and transformer architectures, and static and temporal prediction. Each model was retrained on the common PRECOG dataset while retaining its native input representation, gaze parameterization, and training procedure. Predictions were converted to a common yaw--pitch representation before calibration and evaluation. We additionally implemented two matched custom systems: a custom CNN trained to regress absolute gaze and a differential estimator trained to predict gaze changes.

\textbf{Full-face 448:} The model of Zhang et al.~\cite{zhang2017fullface} directly regresses gaze from a normalized $448 \times 448$ facial image. Its AlexNet-based architecture applies a learned spatial weighting map to the final convolutional features, allowing regions informative of eye and head orientation to contribute selectively to the prediction.

\textbf{RT-GENE}~\cite{fischer2018rtgene} estimates gaze from normalized crops of the left and right eyes together with head pose. The two eye streams are encoded separately before being combined with the pose representation for angular regression, making the contributions of ocular appearance and head orientation explicit.

\textbf{ETH-XGaze.} We use the baseline model~\cite{zhang2020ethxgaze}, which applies a residual convolutional network to a normalized full-face image and predicts gaze pitch and yaw. ETH-XGaze was developed for large variations in camera viewpoint, head pose, and gaze direction, providing a useful contrast to the regime considered here.

\textbf{GazeTR}~\cite{cheng2022gaze} combines convolutional feature extraction with transformer-based self-attention. Its hybrid variant first produces a spatial feature map using ResNet-18 and models global relationships with a transformer encoder before regressing gaze direction.

\textbf{Gaze360}~\cite{kellnhofer2019gaze360} was developed for unconstrained gaze estimation across wide variations in pose, viewing distance, and context. A ResNet-18 encodes a normalized head crop, and the temporal model aggregates consecutive embeddings using a bidirectional LSTM. We evaluate both the single-frame backbone and its temporal model to assess the contribution of temporal context.

\textbf{Direct CNN:} As a baseline control, we trained an ImageNet~\cite{imagenet}-initialized ResNet-18~\cite{he2016deep} on the same binocular crops used by our differential model. The network encodes each crop independently and directly predicts absolute yaw and pitch from a single frame. The differential gaze estimator is built on top of this baseline.

\subsection{Differential Gaze Estimator}

Our custom model of gaze estimation is motivated by the fact that the mapping from eye appearance to gaze direction contains participant- and session-specific terms, including eye shape, camera geometry, and seating position~\cite{park2019few}. All previous models must learn these nuisance factors together with the eye movements of interest. Differential estimation instead predicts the change between two observations, allowing additive recording-specific offsets to cancel from the training target~\cite{liu2019differential}. We adapt that formulation to our setting by encoding two random crops from the same recording with a weight-shared Direct CNN. A two-layer head maps the difference between their embeddings, $e_i$ and $e_j$, to gaze displacement:
\begin{equation}
  \Delta \hat{g}_{ij} = h(e_i - e_j).
\end{equation}
The head has no bias, ensuring $h(0)=0$. At inference, the absolute gaze for query $q$ is reconstructed relative to a labeled anchor $a$:
\begin{equation}
  \hat{g}(q \mid a) = g_a + h(e_q - e_a),
\end{equation}
and predictions are averaged across all anchors. Distributing anchors throughout a recording limits the one-sided temporal extrapolation associated with calibration performed only at session onset. For a temporal variant, we replaced the single crop with a window of 5 frames spanning either 3.2 or 6.4\,s, whose embeddings were contextualized by a one-layer bidirectional GRU~\cite{chung2014empirical} with 128 hidden units; its center step was then passed to the differential head.

\section{Experimental Setup}
\label{sec:setup}

\subsection{Pre-processing}
\label{sec:pre}

Pairing the two modalities required a temporal alignment estimated per recording, since the camera and the eye tracker operated on independent clocks. Linear drift from the camera clock was estimated relative to the nominal 195\,fps rate and then removed, whereas the residual offset was quantified and corrected through blink alignment, allowing a maximum measured discrepancy of 31\,ms. Thirteen participants were excluded because of unreliable EyeLink recordings and 33 because their video and gaze signals could not be reliably aligned. The finalized dataset comprised 140 recordings from 100 participants, totaling approximately 115 hours, or 80.4 million video frames, with an average of 70 minutes per participant.

Facial landmarks were detected using MediaPipe~\cite{lugaresi2019mediapipe} and used to extract the face, head, and eye-region inputs required by each model. The differential and Direct CNN systems received $80 \times 224$ binocular eye crops, while the published models retained their native input regions and resolutions. Where required by the pipeline, inputs were mapped to a normalized camera space using a perspective-warping procedure~\cite{zhang2018revisiting} that standardizes camera distance, scale, and in-plane rotation. EyeLink screen coordinates were projected into three-dimensional camera coordinates using the known display geometry and viewing distance, then expressed as yaw and pitch. Model outputs represented as three-dimensional gaze vectors were converted to the same angular convention.

For evaluation, the 195\,Hz video was first decimated to a common 65\,Hz clip index and then sampled at 5\,Hz for analysis. Frames occurring during EyeLink-annotated blinks were excluded from calibration and evaluation but retained during training. The 100 participants were divided into five subject-disjoint folds, with three folds used for training, one for validation, and one for testing in each rotation. After excluding recordings with fewer than 250 eligible evaluation samples, the common test set comprised 144,743 samples from 105 recordings and 76 participants in total.\vspace{-3mm}

\subsection{Training Protocol}

The differential encoder was initialized from an ImageNet-pretrained ResNet-18 and produced a 256-dimensional embedding. It was trained for 4 epochs from a single seed using AdamW~\cite{loshchilov2017decoupled}, with a batch size of 256, learning rate of $8\times10^{-4}$, weight decay 0.01, and dropout 0.2. The learning rate followed two epochs of linear warm-up and subsequent cosine decay. A smooth-$L_1$ displacement loss was supplemented by a unit-weighted amplitude penalty. The published models were retrained on the same subject partitions using their respective training objectives and implementation procedures. The temporal arms followed the same recipe with the batch reduced to 96 due to memory capacity; the \textit{matched} static arm of Table~\ref{tab:temporal} was retrained under that same schedule to avoid confounds.\vspace{-3mm}

\subsection{Test-time Calibration}

Every model received a common set of 20 labeled calibration frames per recording, distributed approximately uniformly across the valid frames of the session. Each query was required to lie at least 325 frames in the 65\,Hz clip, corresponding nominally to 5\,s, from every anchor. Because anchors were distributed across the full recording and could occur later than the reconstructed query, our primary protocol targets offline reconstruction, or online gaze tracking with assumed distributed calibration phases. For each axis, we fitted an affine correction and clipped its least-squares slope to $[0.5,2.0]$ to prevent unstable scaling from anchors spanning a narrow gaze range.

For the differential model, each anchor was reconstructed using only the other 19 anchors before the affine correction was fitted, because allowing an anchor to reconstruct itself would leak its label into its own prediction. The resulting 20 leave-one-anchor-out (LOO) predictions were paired with their labels to fit the per-axis correction. The published models required no such exclusion because their raw anchor predictions were independent of the labels.\vspace{-3mm}

\subsection{Evaluation protocol}

Frame-level accuracy was measured as the great-circle angle between the predicted and reference gaze directions. We additionally report within-recording Pearson correlations for yaw and pitch. Angular error was evaluated relative to a label-informed constant predictor, i.e., the mean reference gaze of each recording. To assess whether a model follows eye movements rather than remaining near the center of the observed gaze distribution, we also evaluated fine-grained fixation- and saccade-level agreement, where the respective annotations were obtained directly from the EyeLink event stream. Model predictions were averaged within each labeled fixation, while saccades were represented as displacements between two adjacent fixations, retaining only those of at least 1\textdegree\ for analysis. We report the median angular difference between predicted and reference displacement and the within-recording correlation between predicted and reference displacement amplitude. Recordings containing fewer than 20 eligible saccades were excluded from this analysis.

All tables report recording-macro averages of the per-recording metrics. For statistical inference, each participant's differences across their one or two recordings were first averaged. Confidence intervals were then obtained using percentile bootstrap resampling over participant recordings with 10,000 iterations. Pairwise model comparisons used Wilcoxon signed-rank tests, comparisons against the constant predictor used sign-flip tests, whereas FDR correction was used to control for multiple comparisons.\vspace{-2mm}

\section{Results \& Discussion}
\label{sec:results}
\vspace{-1mm}

\subsection{Performance under narrow-range viewing}
\vspace{-1mm}


\begin{table}[t]
\centering
\small
\begin{tabular}{lccc}
\toprule
\textbf{System} & \textbf{Error}\ (\textdegree)\;\textdownarrow & \textbf{Yaw} $r$ & \textbf{Pitch} $r$ \\
\midrule
Full-face 448     & 1.808\,\pmci{.107}\phantom{$^\dagger$} & 0.370\,\pmci{.029} & 0.291\,\pmci{.028} \\
RT-GENE           & 1.784\,\pmci{.107}\phantom{$^\dagger$} & 0.436\,\pmci{.034} & 0.337\,\pmci{.030} \\
ETH-XGaze         & 1.786\,\pmci{.100}\phantom{$^\dagger$} & 0.452\,\pmci{.027} & 0.330\,\pmci{.025} \\
GazeTR            & 1.815\,\pmci{.105}\phantom{$^\dagger$} & 0.490\,\pmci{.028} & 0.314\,\pmci{.023} \\
Gaze360-static    & 1.772\,\pmci{.102}\phantom{$^\dagger$} & 0.525\,\pmci{.027} & 0.414\,\pmci{.024} \\
\midrule
Differential, static & \textbf{1.616}\,\pmci{.098}$^\dagger$ & \textbf{0.636}\,\pmci{.027} & \textbf{0.447}\,\pmci{.030} \\
\bottomrule
\end{tabular}
\vspace{-1mm}
\caption{Recording-macro gaze performance with the 20 anchors and LOO calibration, over 105 sessions from 76 participants. Values are means with the half-width of a 95\% percentile bootstrap interval over recordings, with 10{,}000 resamples. $\dagger$ denotes error significantly below the baseline of constantly predicting the recording mean.}
\label{tab:gaze}
\vspace{-1mm}
\end{table}


\begin{table}[t]
\centering
\small
\setlength{\tabcolsep}{4.8pt}
\begin{tabular}{lccc}
\toprule
\textbf{System} & \textbf{Error}\ (\textdegree)\;\textdownarrow & \textbf{Yaw} $r$ & \textbf{Pitch} $r$ \\
\midrule
Gaze360-LSTM              & 1.698\,\pmci{.098} & 0.546\,\pmci{.029} & 0.455\,\pmci{.024} \\
\midrule
Differential, matched  & 1.626\,\pmci{.098} & 0.640\,\pmci{.025} & 0.428\,\pmci{.033} \\
Differential, 3.2\,s win  & 1.616\,\pmci{.100} & 0.630\,\pmci{.025} & \textbf{0.459}\,\pmci{.030} \\
Differential, 6.4\,s win  & \textbf{1.598}\,\pmci{.098}$^\dagger$ & \textbf{0.646}\,\pmci{.026} & 0.452\,\pmci{.031} \\
\bottomrule
\end{tabular}
\vspace{-1mm}
\caption{Temporal gaze estimation over the same 105 recordings as Table~\ref{tab:gaze}. Values are means with the half-width of a 95\% percentile bootstrap interval over recordings. $\dagger$ denotes error significantly below the static baseline of \textit{Differential-matched}.}
\vspace{-4mm}
\label{tab:temporal}
\end{table}

Table~\ref{tab:gaze} compares the static prediction models under the common 20-anchor protocol. The published estimators remain close to the label-informed recording mean (1.780\textdegree) with error ratios between 0.995 and 1.019, despite exhibiting moderate within-recording correlations with the ground-truth gaze. This effect illustrates the limitation of angular error metrics in evaluating fine-grained gaze estimation in narrow viewing settings. The differential estimator is the only static one to exceed this baseline, reducing angular error to 1.616\textdegree. It also achieves the strongest yaw and pitch correlations, at 0.636 and 0.447, respectively. In paired participant-level comparisons, its angular error is lower and its yaw correlation higher than those of every published comparator after FDR correction (all $p<0.001$). Its improvement over the constant predictor is also significant (sign-flip $p<0.001$), indicating that the reduction in error reflects genuine within-recording gaze tracking.

Temporal context improves both model families, but does not remove the advantage of differential estimation (Table~\ref{tab:temporal}). Gaze360-LSTM, trained on a range of 7 frames, reduces its error ratio to 0.954 to now significantly outperform the constant predictor. Differential estimation, trained on windows of 3.2 and 6.4\,s, performs better at every tested temporal extent, with the 6.4\,s model achieving the lowest angular error and highest yaw correlation, both significantly improved over the matched static counterpart (FDR $p<0.003$). Temporal aggregation thus provides a measurable benefit, whereas differential prediction accounts for most of the observed improvement.\vspace{-2mm}

\begin{figure}[t]
\centering
\includegraphics[width=0.445\textwidth]{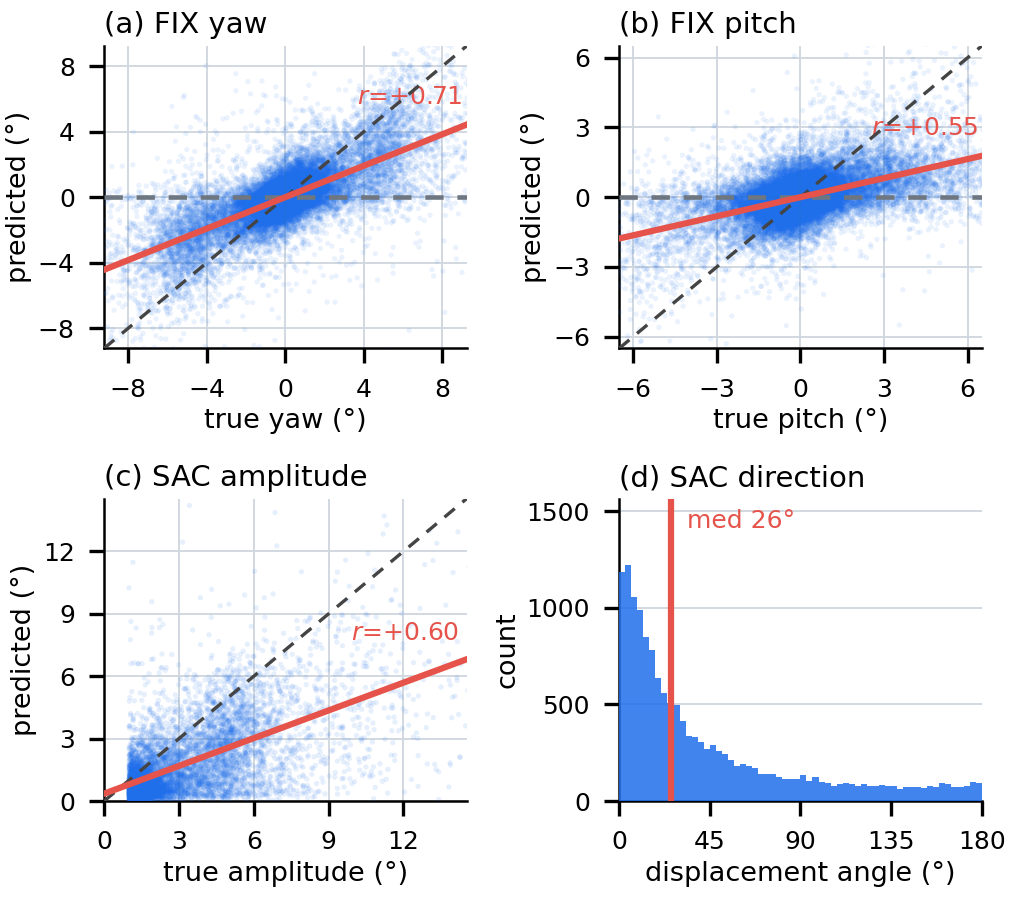}
\vspace{-3mm}
\caption{Event-level analysis of the differential model. \textbf{(a,b)} Predicted fixation position, centered per recording, so the dashed horizontal line is the constant predictor. \textbf{(c)} Predicted against reference saccade amplitude. Saccades of less than 1\textdegree\, were discarded. \textbf{(d)} Distribution of the angle between predicted and reference displacement.}
\label{fig:events}
\vspace{-0.5mm}
\end{figure}

\subsection{Matched-budget anchor comparison}

\begin{table}[t]
\centering
\begin{tabular}{llc}
\toprule
\textbf{System} & \textbf{Anchor use} & \textbf{Error (\textdegree)\;\textdownarrow} \\
\midrule
\multirow{4}{*}{Direct CNN}
    & none & 2.991\,\pmci{.195} \\
    & reconstruction & 2.107\,\pmci{.126} \\
    & affine & 1.894\,\pmci{.114} \\
    & affine, LOO & 1.891\,\pmci{.114} \\
\cmidrule(lr){1-3}
\multirow{2}{*}{Differential}
    & reconstruction & 1.765\,\pmci{.106} \\
    & reconstruction, affine, LOO & \textbf{1.616}\,\pmci{.100} \\
\bottomrule
\end{tabular}
\vspace{-1mm}
\caption{\textbf{Matched-budget anchor benchmark}: every model receives the same 20 frames per session. Recording-macro mean angular error over the 105 recordings for which both absolute and differential systems produce predictions, with the half-width of a 95\% percentile bootstrap interval over recordings.}
\label{tab:matched}
\vspace{-5mm}
\end{table}

Table~\ref{tab:matched} holds the labeled-frame budget, raw-frame identities, data partitions, and held-out participants fixed, while comparing how the two custom pipelines used the same 20 anchors. Without anchors, Direct CNN produced an angular error of 2.991\textdegree. Anchor-based reconstruction reduced it to 2.107\textdegree, while affine calibration lowered it further to 1.894\textdegree. The differential model was more accurate, even without affine correction, reaching 1.765\textdegree\ using anchor-based reconstruction alone and 1.616\textdegree\ after calibration, which corresponds to a participant-level reduction of 0.275\textdegree\ relative to the CNN. 
With front-loaded anchors, average error across all systems increased by 32.1\%.

\subsection{Alignment to event annotations}

Mean angular error can measure absolute proximity to the reference gaze but does not directly establish whether a model followed eye movements or regressed to the session mean. We therefore evaluated video grounding at the level of EyeLink-annotated fixations and saccades. The event-level analysis retained all recordings that contained at least 20 eligible saccades, yielding 69,754 fixations and 15,111 saccades of at least 1\textdegree. Figure~\ref{fig:events} illustrates the (static) differential model's predictions for fixation position, saccade amplitude, and saccade direction. As we show in Table~\ref{tab:events}, the published models differ more significantly on movement-level measures than on frame-level angular error. Gaze360-LSTM is the strongest published comparator, reaching fixation correlations of 0.629 in yaw and 0.533 in pitch, a saccade-amplitude correlation of 0.522, and a median directional error of 35.5\textdegree. The single-frame differential model improves these values to 0.715, 0.548, 0.597, and 31.2\textdegree, respectively. With 6.4\,s of temporal context, it further reduces directional error to 30.7\textdegree. Paired comparisons on both saccade endpoints remain significant against all published configurations (all corrected $p<0.01$).

\begin{table}[t]
\centering
\begin{tabular}{lcccc}
\toprule
& \multicolumn{2}{c}{\textbf{Fixation} $r$} & \multicolumn{2}{c}{\textbf{Saccade}} \\
\cmidrule(lr){2-3}\cmidrule(lr){4-5}
\textbf{System} & \textbf{Yaw} & \textbf{Pitch} & \textbf{Amp.}\ $r$ & \textbf{Dir.}\ (\textdegree) \\
\midrule
Full-face 448     & 0.430 & 0.343 & 0.313 & 48.4 \\
RT-GENE           & 0.517 & 0.413 & 0.415 & 43.5 \\
ETH-XGaze         & 0.541 & 0.404 & 0.415 & 44.4 \\
GazeTR            & 0.570 & 0.389 & 0.434 & 40.4 \\
Gaze360-static    & 0.622 & 0.513 & 0.501 & 39.3 \\
Gaze360-LSTM      & 0.629 & 0.533 & 0.522 & 35.5 \\
\midrule
Differential, static            & 0.715 & \textbf{0.548} & \textbf{0.597} & 31.2 \\
Differential, 6.4\,s win.    & \textbf{0.717} & 0.536 & 0.589 & \textbf{30.7} \\
\bottomrule
\end{tabular}
\caption{\textbf{Event-level agreement} over annotated fixations and saccades, averaged across the 103 recordings with at least 20 saccades. Saccade columns give the within-recording correlation of predicted with reference amplitude and the median angle between predicted and reference displacement. Bootstrap 95\% half-widths are near $\pm$0.03 on the correlations and $\pm$3\textdegree\ on direction for every system.}
\label{tab:events}
\vspace{-4mm}
\end{table}

Frame-level and movement-level evaluation therefore capture different aspects of model sensitivity. Across the published configurations, the rankings produced by angular error and saccade agreement were only weakly aligned. GazeTR, for example, had the largest frame-level angular error but substantially better directional agreement than that error would suggest for other backbones. Reporting movement-level agreement alongside angular error therefore distinguishes models that track the eye from those that merely drift to the center of its observed range.\vspace{-1mm}

\section{Conclusion}
\label{sec:conc}

We evaluated appearance-based gaze estimation under narrow-range screen viewing using a subject-disjoint protocol. Published models remained close to a recording-mean predictor, whereas differential estimation achieved the lowest angular error and strongest fixation- and saccade-level agreement, with further gains from provided temporal context. Relative to the uncalibrated direct regressor, the complete differential pipeline reduced error by 46\% in total. These findings support differential estimation for reliable gaze reconstruction in constrained viewing settings. Generalization remains nevertheless limited by the college-aged cohort of our dataset, the single camera geometry of our setup, and the offline calibration.

\section{Compliance with Ethical Standards}
This work is an analysis of previously collected human-subject data. The original study received institutional review board approval, and all participants provided written informed consent, as described in~\cite{precog-methods}. No new data were collected for this work.

\bibliographystyle{IEEEbib}
\bibliography{refs}

\end{document}